\documentclass[sigconf, nonacm]{acmart}

\usepackage{booktabs} 
\usepackage{hyperref}
\usepackage{url}
\usepackage{graphicx}
\usepackage{caption,subcaption}
\usepackage{amsmath,amssymb}

\DeclareMathOperator*{\argmax}{arg\,max}

\copyrightyear{2019}
\acmYear{2019} 
\setcopyright{iw3c2w3}
\acmConference[WWW '19]{Proceedings of the 2019 World Wide Web Conference}{May 13--17, 2019}{San Francisco, CA, USA}
\acmBooktitle{Proceedings of the 2019 World Wide Web Conference (WWW '19), May 13--17, 2019, San Francisco, CA, USA}
\acmPrice{}
\acmDOI{10.1145/3308558.3313416}
\acmISBN{978-1-4503-6674-8/19/05}

\usepackage{balance}

\begin{document}
\title[Adversarial Sampling and Training for Semi-Supervised Information Retrieval]{Adversarial Sampling and Training for \\ Semi-Supervised Information Retrieval}



\author{Dae Hoon Park}
\affiliation{%
 \institution{Huawei Research America}
 \state{California, United States}
}
\email{dae.hoon.park@huawei.com}

\author{Yi Chang}
\affiliation{%
  \institution{School of Artificial Intelligence, Jilin University}
  \institution{Key Laboratory of Symbolic Computation and Knowledge Engineering of Ministry of Education}
  \city{China}
}
\email{yichang@jlu.edu.cn}


\begin{abstract}
Ad-hoc retrieval models with implicit feedback often have problems, \textit{e.g.}, the imbalanced classes in the data set.
Too few clicked documents may hurt generalization ability of the models, whereas too many non-clicked documents may harm effectiveness of the models and efficiency of training.
In addition, recent neural network-based models are vulnerable to adversarial examples due to the linear nature in them.
To solve the problems at the same time, we propose an adversarial sampling and training framework to learn ad-hoc retrieval models with implicit feedback.
Our key idea is (\texttt{i}) to augment clicked examples by adversarial training for better generalization and (\texttt{ii}) to obtain very informational non-clicked examples by adversarial sampling and training.
Experiments are performed on benchmark data sets for common ad-hoc retrieval tasks such as Web search, item recommendation, and question answering.
Experimental results indicate that the proposed approaches significantly outperform strong baselines especially for high-ranked documents, and they outperform IRGAN in NDCG@5 using only 5\% of labeled data for the Web search task.

\end{abstract}



\maketitle

\section{Introduction}\label{sec:introduction}
Ad-hoc retrieval systems provide a ranked list of documents given a query, in which a user's information need is expressed.
Such ad-hoc retrieval systems prevail in our daily lives, from Web search and question answering to product recommendation, which can be regarded as ad-hoc retrieval with users being queries and products being documents.
Modern ad-hoc retrieval models are usually supervised or semi-supervised with training data to learn the relevance of documents to a query, because of its outstanding performance over unsupervised models.

Generally speaking, the training data are obtained for (semi-)supervised ad-hoc retrieval models in two ways.
Annotators can explicitly label documents with their relevance to a query, and such labels are called explicit feedback.
It is, however, often too expensive to obtain enough explicit feedback, and the labels may contain bias from a few annotators. 
Instead, labels can be inferred from the user's reactions (\textit{e.g.}, clicks and views) on the given documents, and such reactions are called implicit feedback.
Implicit feedback is cheap and reflects preference of actual users, so it has been widely studied \cite {joachims2005accurately, shen2005context, radlinski2005query} and adopted in the industry.
Typically, there are much more non-clicked documents than clicked ones.
Too few clicked documents can be problematic for generalization of the learned models.
On the other hand, too many non-clicked documents can slow down the training and harm effectiveness of the trained model.
Especially for ad-hoc retrieval tasks, the majority of the non-clicked documents is often not informational, sampling informational non-clicked documents is important for efficiency and effectiveness \cite{rendle2014improving, zhang2013optimizing}.

Meanwhile, ad-hoc retrieval models often have a problem in themselves; their structures are vulnerable to adversarial examples.
That is, due to the linearity in many models, the output can be dramatically changed by infinitesimal changes in input dimensions \cite{goodfellow2014generative}.
Traditional models such as logistic regression and matrix factorization work in a very linear way, and even recent deep neural networks are also designed to work in a quite linear way \cite{goodfellow2014explaining} (\textit{e.g.}, ReLU \cite{glorot2011deep} and LSTMs \cite{hochreiter1997long}).

To solve the aforementioned problems in implicit feedback and ad-hoc retrieval models, we propose an adversarial sampling-based adversarial training framework.
On one hand, we generate adversarial positive examples to augment labeled (or clicked) documents with informational ones.
On the other hand, to obtain informational negative examples, we first sample difficult examples adversarially from unlabeled (or non-clicked) documents.
Then, we further generate adversarial negative examples, which are even more informational, on top of the sampled negative examples.
The generated adversarial examples are supposed to be informational and to remove the weakness in the model's linearity.
We also propose virtual adversarial training-based approach, which does not require labels to generate adversarial examples and thus suitable for semi-supervised learning, and its variant that is more efficient and effective.

We perform experiments on benchmark data sets for three popular tasks: Web search, item recommendation, and question answering.
Our proposed approaches are mainly compared with a strong baseline, IRGAN \cite{wang2017irgan}, which adopts a different kind of adversarial training to sample negative documents.
Experimental results indicate that our proposed approaches are very effective on ad-hoc retrieval tasks and significantly outperform baselines especially on Web search and item recommendation.
Moreover, our proposed approaches outperform IRGAN in NDCG@5 with only 5\% of the labeled data for Web search.

Our contributions in this paper include:
\begin{itemize}
    \item We propose a novel framework for learning ad-hoc retrieval models with implicit feedback. We generate very informational training examples by adversarial sampling and training. To the best of our knowledge, there has been no research work that incorporates adversarial sampling with adversarial training for ad-hoc retrieval models.
    \item We also propose virtual adversarial training and its variation for ad-hoc retrieval models. They can generate adversarial examples without labels, which may be ideal for noisy implicit feedback.
    \item We perform experiments on benchmark data sets for popular tasks. Experimental results indicate the proposed approaches significantly outperform strong baselines. The proposed approaches are empirically shown to be data-efficient.
\end{itemize}

\section{Related Work}
\subsection{Adversarial Training}
It was found in \cite{szegedy2013intriguing} that several machine learning models, including modern state-of-the-art neural networks, fail with adversarial examples.
A slight adversarial perturbation in the original input was enough to fake the models, meaning that the models classify the adversarial example into a wrong class with high confidence.
Then, Goodfellow \textit{et al.} hypothesized that the vulnerability of the models come from the linear nature in the models \cite{goodfellow2014explaining}.
That is, the adversarial perturbation in each dimension adds up to a great change in the output due to some extent of linearity in the models.
Goodfellow \textit{et al.} also suggested to train a model by learning adversarial examples as well as original examples, where the adversarial examples were generated with the proposed \textit{fast gradient sign method}.
The trained model was reported more robust to adversarial examples.
Adversarial examples and adversarial training are surveyed in \cite{warde201611}.
Meanwhile, the earlier adversarial training approaches required labels of examples, and this is not ideal for semi-supervised learning where there are much more unlabeled data than labeled data.
Miyato \textit{et al.} \cite{miyato2015distributional} proposed virtual adversarial training, where labels are not required to generate adversarial examples. 
With its strength in semi-supervised learning, it has been applied to semi-supervised text classification \cite{miyato2016adversarial} and image classification \cite{miyato2018virtual}.

We adopt the ideas of adversarial training and virtual adversarial training in our approaches.
However, there are several differences between our approaches and them.
In our task, we assume that training data in implicit feedback consist of relatively fewer single-class labeled data and much more unlabeled data, where there exist many non-informational unlabeled examples.
Such informativeness of unlabeled examples is not studied in the original works.
We pay attention to it in this paper, especially for ad-hoc retrieval with implicit feedback.
We further combine adversarial sampling with adversarial training to boost the informativeness of unlabeled examples.
Meanwhile, virtual adversarial training iterates over all unlabeled examples and thus may not be efficient and effective for implicit feedback.
Hence, we propose selective virtual adversarial training that iterates over only labeled examples and adversarially sampled unlabeled examples.
Our problem, ad-hoc retrieval, also makes our approaches different from original ones.
In our problem, there can be two input variables, queries and documents, instead of one single variable.
In addition, we apply adversarial training to a pair-wise learning-to-rank framework.

\subsection{Adversarial Training for Ad-hoc Retrieval}
Recently, deep neural networks-based approaches have been successfully adopted in information retrieval tasks such as Web search \cite{huang2013learning, guo2016deep}, click modeling \cite{borisov2016neural}, and query suggestion and auto-completion  \cite{sordoni2015hierarchical, park2017neural}.
Adversarial training has recently gained popularity with neural networks, and a few adversarial training-based approaches have been proposed for ad-hoc retrieval tasks.
For example, an idea of generative adversarial networks (GAN) \cite{goodfellow2014generative} has been adopted to unify generative and discriminative information retrieval models by IRGAN \cite{wang2017irgan}.
Its goal is indeed similar to ours, which is to build effective ad-hoc retrieval models through difficult examples.
The generator of IRGAN tries to fool the discriminator by providing adversarial negative examples while the discriminator tries to distinguish them from true positive examples so that the generator and the discriminator can mutually enhance.
Through dynamically sampling more and more difficult examples by the evolving generator, IRGAN achieves outstanding performance.
However, IRGAN is different from our approaches in many aspects.
Although it contains a generator, IRGAN does not really generate negative examples but samples them from unlabeled data according to the generator model.
On the other hand, our approach \textit{generates} difficult examples on top of existing examples by adding adversarial perturbation to them.
Incorporating adversarial sampling with adversarial training, our approaches can generate even more difficult negative examples based on the adversarially sampled negative examples.
Also, our approaches require one single model whereas IRGAN consists of two models: a generator and a discriminator; this means the required number of parameters and hyperparameters can double, and training both models at the same time can be more difficult.
Furthermore, IRGAN requires a pre-trained model to ensure stability during training, unlike our stand-alone model.
\cite{yang2018adversarial} extends IRGAN with a different neural network architecture for question answering.

More recently, a simultaneous work but independent from ours has proposed adversarial training for item recommendation \cite{he2018adversarial}.
It employs adversarial perturbation to build a robust model, and the experiment results show that it is superior to existing models.
Despite its novel approach, our work is different from it in several aspects.
Their approach adds adversarial perturbation to the model parameters while we add it to the input.
Also, they do not have the same assumption on data as ours, \textit{i.e.}, single-class labeled data and much more unlabeled data, so they do not focus on sampling difficult negative examples from unlabeled data.
Their focus is on item recommendation, but our focus is on ad-hoc retrieval with implicit feedback.
Lastly, we further explore virtual adversarial training that can be promising for semi-supervised learning, which is not in their scope.

\subsection{Negative Sampling for Ad-hoc Retrieval}
Negative example sampling techniques have been adopted for information retrieval tasks as well as natural language processing tasks.
Word embedding models have sampled negative examples by their frequency \cite{mikolov2013distributed,grbovic2015context}, and \cite{rao2016noise} employs ``max sampling'' that samples negative examples that are most similar to positive examples for question answering.
Generative Adversarial Networks-based negative sampling also has been proposed for ad-hoc retrieval tasks \cite{wang2017irgan}.

Negative sampling techniques have been studied more extensively for item recommendation.
To train a model efficiently, negative sampling techniques have been employed.
Traditional models including Bayesian Personalized Ranking \cite{rendle2009bpr} rely on uniform sampling for negative examples.
Dynamic negative sampling techniques \cite{rendle2014improving, zhang2013optimizing} that sample informational negative examples for the current model also have been proposed.
Recently, negative sampling by leveraging view information has been proposed and empirically shown to be effective for e-commerce data sets \cite{ding2018improved}.
Hidasi and Karatzoglou \cite{hidasi2018recurrent} recently proposed a BPR-max loss function that assigns larger weights to more informative negative examples.
To the best of our knowledge, no previous ad-hoc retrieval models adversarially generate negative examples on top of sampled negative examples in order to generate even more informational examples, as our approaches do.

\section{Problem Definition}
We study a typical ad-hoc retrieval problem.
Given a query $q$ that contains a user's information need and a set of documents $D=\{d_i\}_{i=1}^{M}$, the goal is to retrieve and rank documents based on their estimated relevance to $q$, where Probability Ranking Principle is assumed.
Ad-hoc retrieval problems are not limited to document search, and $q$ and $D$ may be in other types.
Item recommendation can be regarded as an ad-hoc retrieval problem where $q$ is a user and $D$ is a set of items.
Question answering (retrieval-based) also can fall within ad-hoc retrieval where $q$ is a question and $D$ is a set of answer candidates.

We specifically study semi-supervised ad-hoc retrieval with implicit feedback, where for each $q$, we have a set of labeled (clicked) documents, which we assume relevant, and unlabeled (non-clicked) documents.
We also assume the number of labeled documents is much less than that of unlabeled documents.
Indeed, such configuration is common in the industry.
When the retrieved documents are shown to users, the users usually click only a few of them while many documents are not clicked.
The non-clicked documents are not necessarily irrelevant to $q$, so they are left unlabeled or sometimes labeled as \textit{viewed}.

The labeled and unlabeled documents for $q$ are given to a model as training data, whose example is in a form of triple $(q, d, y)$, where $y$ is a relevance of $d$ to $q$, and $y=1$ if a user clicked $d$ for $q$.
Non-relevant documents ($y=0$) come from unlabeled documents.
The ad-hoc retrieval model learns a function $f_{\boldsymbol{\theta}}: (q,d) \mapsto y$, where ${\boldsymbol{\theta}}$ is a set of model parameters.
The learned function $f_{\boldsymbol{\theta}}$ is then employed to predict a relevance score of a test ($q$, $d$) pair.

There are challenges to learn ad-hoc retrieval models from implicit feedback.
If the number of clicked documents is small, it can easily suffer from over-fitting.
That means, the model will not generalize well on the test data. 
On the other hand, there are relatively too many unlabeled documents, and many of the unlabeled documents are either redundant or obviously irrelevant.
That is, the unlabeled documents are not informational so that the model will not learn effectively with them.
Therefore, careful utilization of the imbalanced data is desired to build an effective ad-hoc retrieval model.

\section{Adversarial Sampling and Training Framework for Ad-hoc Retrieval}

In order to address the challenges in implicit feedback, we propose a framework of adversarial sampling and training that are applied differently for labeled and unlabeled documents.
On one hand, we employ adversarial training to build a generalizable model with relatively few labeled documents.
On the other hand, we first adversarially sample informational examples, and on top of them, we further generate even more informational examples by adversarial training.
The generated adversarial examples for both labeled and unlabeled documents also help ad-hoc retrieval models cope with the weakness in linearity of the models.

In this section, we first explain how adversarial examples can cause a significant change in the output of models that have linearity.
We then propose multiple adversarial training methods that can generate informational examples based on existing training examples.
Then, adversarial sampling is incorporated with adversarial training in order to amplify informativeness of unlabeled examples.
Pairwise learning is also proposed to effectively learn with implicit feedback.

\subsection{Adversarial Examples for Ad-hoc Retrieval}\label{sec:adv_examples}
An adversarial example is defined as an example that is slightly perturbed from the original example but greatly changes the activation and thus the output.
Adversarial examples were first introduced for neural networks in \cite{szegedy2013intriguing} and later found that they occur due to linearity in models \cite{goodfellow2014explaining}.
For an adversarial input $\mathbf{\tilde{x}}=\mathbf{x}+\boldsymbol{\eta}$, where $\mathbf{x}$ is an original input vector and $\boldsymbol{\eta}$ is an adversarial perturbation vector of the same shape, 
a dot product between $\mathbf{\tilde{x}}$ and a weight vector $\mathbf{w}$ becomes
\begin{equation}
    \begin{aligned}
        \mathbf{w}^\intercal \mathbf{\tilde{x}} = \mathbf{w}^\intercal \mathbf{x} + \mathbf{w}^\intercal \boldsymbol{\eta} . \nonumber
    \end{aligned}
\end{equation}
The change of the activation by adversarial perturbation is thus $\mathbf{w}^\intercal \boldsymbol{\eta}$.
If the average magnitude of $\mathbf{w}$ and $\boldsymbol{\eta}$ are $m$ and $\epsilon$, respectively, then the maximum change caused by perturbation can grow linearly by $\epsilon m n$, where $n$ is the number of dimensions in $\mathbf{x}$.
(We explain how to approximate the optimal $\boldsymbol{\eta}$ in the next section.)
Hence, small changes by perturbation can accumulate with dimensions to a great change.
Such a great change can propagate to an output of more complicated models.
For example, neural network activation functions such as ReLUs \cite{glorot2011deep} and sigmoid, which are widely used in modern neural networks, are piece-wise linear or almost linear (in non-saturating section); hence, the great change can easily propagate to the final output \cite{goodfellow2014explaining}.

Such perturbations can also exist in ad-hoc retrieval systems.
For example, let's assume $\mathbf{x}$ and $\mathbf{x'}$ are topic probabilities estimated by topic models for two semantically identical documents, where only $\mathbf{x}$ is in the training data.
A slight word change from $\mathbf{x}$ to $\mathbf{x'}$ can cause slight difference in each dimension of $\mathbf{x}$ and $\mathbf{x'}$, and this may result in a large change in the output, yielding wrong prediction on $\mathbf{x'}$.
Similarly, less processed features such as raw text and TF-IDF values can also cause such effect.
For example, two semantically identical documents may have different raw text features, by synonyms of their words.
Although their raw features are different, their latent vectors may still be similar, but the subtle difference in the latent vectors may eventually result in a great change in the output.
This is because the latent vectors still need to go through activation functions that are linear or almost linear.

Unfortunately, recent ad-hoc retrieval models as well as several traditional models are often vulnerable to adversarial examples.
Neural networks are recently employed for state-of-the-art performance, and modern neural networks behave in a linear way (\textit{e.g.}, ReLUs) for more effective learning by avoiding vanishing gradients.
Earlier models such as RankNet \cite{burges2005learning}, or matrix factorization \cite{koren2009matrix} also operate in a quite linear way.

\subsection{Adversarial Training for Robust Ad-hoc Retrieval Models}\label{sec:adv_training}

\subsubsection{Adversarial Training}
We propose to train ad-hoc retrieval models with adversarial examples that can locate the weak points in the models.
In general, ad-hoc retrieval models are trained by minimizing the following cost function
\begin{equation}\label{eq:general}
    \begin{aligned}
        \mathbb{E}_{q,d,y \sim p_{\text{data}}} \mathrm{J}(\mathbf{q}, \mathbf{d}, y; \boldsymbol{\theta})
    \end{aligned}
\end{equation}
where $\mathbf{q}$ and $\mathbf{d}$ are feature vectors for $q$ and $d$, respectively, and $y \in \{0,1\}$ is a relevance label, and $\boldsymbol{\theta}$ is a set of model parameters.\footnote{If the given data consist of one input vector $\mathbf{x}$ instead of two vectors $\mathbf{q}$ and $\mathbf{d}$ for a query and a document as in Section \ref{sec:adv_examples}, $\mathbf{x}$ can replace $\mathbf{q}$ and $\mathbf{d}$ accordingly. }
$\mathrm{J}$ can be, for example, a cross entropy loss function.
Hence, the goal is to learn a function $f_{\boldsymbol{\theta}}: (\mathbf{q},\mathbf{d}) \mapsto y$.
Intuitively, in order to build a model that is robust to adversarial perturbation, we can generate adversarial perturbation and let the model learn from adversarial examples that are generated by adding adversarial perturbation to original training examples.
That is, the adversarial examples still keep the original labels, but their $\mathbf{q}$ and $\mathbf{d}$ are modified to become more difficult (\textit{i.e.}, yield greater losses).
Learning with more difficult examples that attack the model's weaknesses, the model can become more robust to adversarial perturbation.
Therefore, we can add a cost for adversarial examples to the objective as follows:
\begin{equation}\label{eq:general_objective}
    \begin{aligned}
        \mathbb{E}_{q,d,y \sim p_{\text{data}}}
            \mathrm{J}(\mathbf{q}, \mathbf{d}, y; \boldsymbol{\theta}) +
            \alpha \mathrm{J}(\mathbf{q}+\boldsymbol{\eta}_q, \mathbf{d}+\boldsymbol{\eta}_d, y; \boldsymbol{\theta})
    \end{aligned}
\end{equation}
where $\boldsymbol{\eta}_q$ and $\boldsymbol{\eta}_d$ are adversarial perturbations for $\mathbf{q}$ and $\mathbf{d}$, respectively, and $\alpha$ is a hyperparameter, which is set to 1 in this work.
This objective means that regardless of adversarial perturbation, the model should learn to predict the same relevance.

\subsubsection{Generating Adversarial Perturbation}
When generating adversarial perturbation, the magnitudes of perturbation vectors need to be limited so that the adversarial examples do not become too similar to examples of the opposite classes.
Thus, their magnitudes are limited by a hyperparameter $\epsilon$ such that $||\boldsymbol{\eta}||_p < \epsilon$.
In order to build a model that is robust to perturbation, we need to generate a perturbation vector that can achieve the greatest loss within the limit of $\epsilon$.
In other words, a model trained with more difficult examples can be more effective for the unseen test examples.
Such perturbation vectors are defined as
\begin{equation}\label{eq:general_perturbation}
    \begin{aligned}
        \boldsymbol{\eta}_q, \boldsymbol{\eta}_d = 
        \argmax_{\boldsymbol{\eta}_q: ||\boldsymbol{\eta}_q||_p < \epsilon, \,
        \boldsymbol{\eta}_d: ||\boldsymbol{\eta}_d||_p < \epsilon }
        \mathrm{J}(\mathbf{q}+\boldsymbol{\eta}_q, \mathbf{d}+\boldsymbol{\eta}_d, y; \boldsymbol{\hat{\theta}}) .
    \end{aligned}
\end{equation}
We use the same $\epsilon$ for $\boldsymbol{\eta}_q$ and $\boldsymbol{\eta}_d$ in this paper, but different values can be used.
Also, $\boldsymbol{\hat{\theta}}$ denotes a copy of $\boldsymbol{\theta}$, in order to avoid propagating gradients from this perturbation generation process to $\boldsymbol{\theta}$.
Note that the objective in \eqref{eq:general_objective} with the equation \eqref{eq:general_perturbation} can be interpreted as a \textit{minimax} game.
Similar to the \textit{fast gradient sign method} \cite{goodfellow2014explaining}, with first-order Taylor series approximation, equation \eqref{eq:general_perturbation} is approximated as
\begin{equation}\label{eq:general_taylor}
    \begin{aligned}
        \boldsymbol{\eta}_q = &
        \argmax_{\boldsymbol{\eta}_q: ||\boldsymbol{\eta}_q||_p < \epsilon }
        \mathrm{J}(\mathbf{q}, \mathbf{d}, y; \boldsymbol{\hat{\theta}}) 
        + \boldsymbol{\eta}_q \nabla_{\mathbf{q}}  \mathrm{J}(\mathbf{q}, \mathbf{d}, y; \boldsymbol{\hat{\theta}})   \\
        \boldsymbol{\eta}_d = &
        \argmax_{ \boldsymbol{\eta}_d: ||\boldsymbol{\eta}_d||_p < \epsilon }
        \mathrm{J}(\mathbf{q}, \mathbf{d}, y; \boldsymbol{\hat{\theta}}) 
        + \boldsymbol{\eta}_d \nabla_{\mathbf{d}}  \mathrm{J}(\mathbf{q}, \mathbf{d}, y; \boldsymbol{\hat{\theta}}) .
    \end{aligned}
\end{equation}
Its solution depends on the value of $p$ in the $p$-norm as follows
\begin{equation}\label{eq:general_eta}
    \begin{aligned}
        \boldsymbol{\eta}_q = 
        \begin{cases}
        \epsilon \text{sign}(\mathbf{g}_q) \,\, & \text{if} \,\, p=\infty \\
        \epsilon \frac{\mathbf{g}_q}{||\mathbf{g}_q||_2} & \text{if} \,\, p=2
        \end{cases}
        \,\, \text{where} \,\, \mathbf{g}_q = \nabla_{\mathbf{q}}  \mathrm{J}(\mathbf{q}, \mathbf{d}, y; \boldsymbol{\hat{\theta}})
    \end{aligned}
\end{equation}
where $\mathbf{g}_q$ can be efficiently computed by backpropagation while solving the first term of \eqref{eq:general_objective}.
That means, adversarial training requires only a few additional computations that can be done efficiently.
$\boldsymbol{\eta}_d$ can be solved in the same way, so we do not include its solution here.
Regarding the choice of $p$, a max norm ($p=\infty$) is used in \cite{goodfellow2014explaining} because the magnitudes of perturbation in image pixels are supposed to be small.
However, we do not have such assumption for text; indeed, semantically similar documents can have very different text at least in their raw representation.
Also, L2 norm can sometimes perform better than max norm for adversarial training  \cite{miyato2018virtual}, so we employ L2 norm in this work.

\subsubsection{Virtual Adversarial Examples}
The process for generating adversarial examples requires labels of query-document pairs.
However, the number of unlabeled (non-clicked) documents is usually much greater than that of labeled (clicked) documents in implicit feedback.
Even if we regard unlabeled ones as negative examples, the generated adversarial examples from the negative examples may not be ideal due to the uncertainty in the assumed labels.
Virtual Adversarial Examples \cite{miyato2015distributional} may thus be useful in this case, which do not require labels to generate adversarial examples.
The following cost function $\mathrm{J_{KL}}$ is added to the objective in \eqref{eq:general} for both labeled and unlabeled data:
\small
\begin{equation}\label{eq:vat}
    \begin{aligned}
        &\mathrm{J_{KL}}(\mathbf{q}, \mathbf{d};\boldsymbol{\theta}) = 
        \text{KL} \big[ p(\cdot|\mathbf{q},\mathbf{d};\boldsymbol{\theta}) || p(\cdot|\mathbf{q}+\boldsymbol{\eta}_q,\mathbf{d}+\boldsymbol{\eta}_d;\boldsymbol{\theta})  \big]  \,\, \text{, where} \\
        &\boldsymbol{\eta}_q, \boldsymbol{\eta}_d = 
        \argmax_{\boldsymbol{\eta}_q: ||\boldsymbol{\eta}_q|| < \epsilon, \,
        \boldsymbol{\eta}_d: ||\boldsymbol{\eta}_d|| < \epsilon } 
        \text{KL} \big[ p(\cdot|\mathbf{q},\mathbf{d};\boldsymbol{\hat{\theta}}) || p(\cdot|\mathbf{q}+\boldsymbol{\eta}_q,\mathbf{d}+\boldsymbol{\eta}_d;\boldsymbol{\hat{\theta}})  \big]
    \end{aligned}
\end{equation}
\normalsize
where $\text{KL}\big[ p || p' \big]$ denotes KL divergence between conditional relevance distributions $p$ and $p'$.
Basically, minimizing the objective including this cost function means that we want to reduce the distribution difference that is caused by adversarial perturbation.
In other words, we want to enhance the model's local smoothness of conditional relevance distribution so that the model's output does not dramatically change by adversarial perturbation.
The perturbation vector can be computed by a second-order Taylor series approximation and a single iteration of power method on the KL divergence function as in \cite{miyato2015distributional}.
The solution thus can be approximated as
\begin{equation}\label{eq:vat_perturbation}
    \begin{aligned}
        \boldsymbol{\eta}_q &= 
        \epsilon \frac{\mathbf{g}_q}{||\mathbf{g}_q||_2} 
        \,\, \text{,} \,\, \boldsymbol{\eta}_d = 
        \epsilon \frac{\mathbf{g}_d}{||\mathbf{g}_d||_2} \text{, where} \\
        \mathbf{g}_q &= \nabla_{\mathbf{q}+\mathbf{e}_q}  
        \text{KL} \big[ p(\cdot|\mathbf{q},\mathbf{d};\boldsymbol{\hat{\theta}}) || p(\cdot|\mathbf{q}+\mathbf{e}_q,\mathbf{d};\boldsymbol{\hat{\theta}})  \big] \\
        \mathbf{g}_d &= \nabla_{\mathbf{d}+\mathbf{e}_d}  
        \text{KL} \big[ p(\cdot|\mathbf{q},\mathbf{d};\boldsymbol{\hat{\theta}}) || p(\cdot|\mathbf{q},\mathbf{d}+\mathbf{e}_d;\boldsymbol{\hat{\theta}})  \big]
    \end{aligned}
\end{equation}
where $\mathbf{e}_q$ and $\mathbf{e}_d$ are small random vectors.
This process does not require labels, so the uncertainty of relevance in implicit feedback is not a problem to generate perturbations, and thus all non-clicked documents can be safely used for semi-supervised learning.
However, training with all non-clicked documents can be very inefficient, so we discuss sampling approaches in Section \ref{sec:sampling}.

\subsubsection{Adversarial Examples for Discrete Input}
If an input is a discrete value such as a word or an item ID, where the value is converted to a latent vector, one may consider to add a perturbation vector to the latent vector (or embedding vector).
However, adding it to the perturbation vectors may be tricky because they may learn to increase their magnitudes so that the amount of perturbation $\epsilon$ becomes negligible.
To avoid such phenomenon, it was proposed in \cite{miyato2016adversarial} to normalize the latent vector after adding a perturbation vector to it.
Meanwhile, ad-hoc retrieval models can have continuous input as well as discrete one, \textit{e.g.,} TF-IDF features and other pre-processed features.
To serve both discrete and continuous input, we employ another trick that adds adversarial perturbation to the input.
When the input is continuous, we can add the perturbation vector as usual: $\mathbf{\tilde{x}} = \mathbf{x}+\boldsymbol{\eta}$.
For discrete input, we are given $\mathbf{x}$ that is a length-$v$ one-hot encoded input vector, where $v$ is the cardinality.
Instead of looking up the length-$k$ embedding vector $\mathbf{z} \in \mathbb{R}^k$ from the embedding matrix $\mathbf{Z} \in \mathbb{R}^{v\times k}$, we multiply $\mathbf{\tilde{x}}$ and $\mathbf{Z}$ together to obtain the perturbed embedding vector of $\mathbf{x}$.
That is, the perturbed embedding vector $\mathbf{\tilde{z}}$ is obtained by
\begin{equation}\label{eq:embedding}
    \begin{aligned}
        \mathbf{\tilde{z}} = (\mathbf{x}+\boldsymbol{\eta}) \mathbf{Z}.
    \end{aligned}
\end{equation}
That means, for example, we make a mixture of words to generate a perturbed embedding vector for the input word.
The advantage of this approach is that it can be extended to accommodate the input that is combination of discrete values and continuous values.
In addition, this approach may potentially provide more interpretable perturbation as it is engaged to the original input.
For example, a perturbation by ``\texttt{- man + woman}'' for the input ``\texttt{king}'' will indicate that the perturbation changes the gender to generate the embedding vector of ``\texttt{queen}''.
Such interpretability may be interesting since that of neural networks-based classifiers has been weak and has attracted attention  \cite{zhang2017interpretable, xu2018interpreting}.
We do not explore these advantages in this paper as they are out of its scope, but they are left for our future work.

\subsection{Adversarial Training with Adversarial Sampling} \label{sec:sampling}
Adversarial training can be interpreted as learning with difficult examples that are generated to attack the current model's weakness.
To amplify the effectiveness, we propose adversarial sampling-based adversarial training that generates even more difficult examples from already difficult examples.
In ad-hoc retrieval systems with implicit feedback, labeled documents are mainly relevant but the number of documents is relatively small.
On the other hand, unlabeled documents are more likely to be negative than positive while the number of documents is much greater.
As many of the unlabeled examples may be either redundant or obviously irrelevant, finding informational examples may be the key to effective and efficient model.

Therefore, we use existing labeled documents as positive examples, but we adversarially sample negative examples from unlabeled documents.
The optimization objective for adversarial training is thus defined as
\begin{equation}\label{eq:adv_training}
    \begin{aligned}
        \sum_q \Big( & \mathbb{E}_{d \sim p_{\text{data}}(d|q,y=1)} \big[ \mathrm{J}(\mathbf{q}, \mathbf{d}, y=1; \boldsymbol{\theta}) +
            \mathrm{J}(\mathbf{q}+\boldsymbol{\eta}_q, \mathbf{d}+\boldsymbol{\eta}_d, y=1; \boldsymbol{\theta}) \big] + \\
        & \mathbb{E}_{d \sim p_{\boldsymbol{\theta}}(d|q,y=1) } \big[ \mathrm{J}(\mathbf{q}, \mathbf{d}, y=0; \boldsymbol{\theta}) +
            \mathrm{J}(\mathbf{q}+\boldsymbol{\eta}_q, \mathbf{d}+\boldsymbol{\eta}_{d}, y=0; \boldsymbol{\theta}) \big]
        \Big)
    \end{aligned}
\end{equation}
where $\mathbf{q}$ and $\mathbf{d}$ are feature vectors for $q$ and $d$, respectively, and the adversarial perturbation vectors $\boldsymbol{\eta}_q$ and $\boldsymbol{\eta}_d$ are computed as in Section \ref{sec:adv_training} for each example.
All positive examples are selected from labeled data by $p_\text{data}$, whose distribution is uniform, and the negative examples are selected from unlabeled data by $p_\mathbf{\theta}$.
In practice, the model goes through all positive examples in the labeled data while it goes through the same number of negative examples that are sampled from unlabeled data.
Negative examples are sampled by $p_{\boldsymbol{\theta}}(d|q,y=1)$ to ensure the examples are adversarial ($y=1$) so that they are difficult for the current model.
Note that the negative examples are sampled dynamically for the current model, and more efficient sampling can be done by estimating $p_{\boldsymbol{\theta}}(d|q,y=1)$ for every $K$ epochs or estimating it for only $C$ document candidates.
The conditional probability for sampling $d$ can be estimated by the following softmax function:
\begin{equation}
    \begin{aligned}
        p_{\boldsymbol{\theta}}(d|q,y=1) = 
            \frac{ \exp (f_{\boldsymbol{\theta}} (\mathbf{q}, \mathbf{d}) / \tau ) }
            { \sum_{d'} \exp (f_{\boldsymbol{\theta}} (\mathbf{q}, \mathbf{d}') / \tau) }
    \end{aligned}
\end{equation}
where $\tau$ is a temperature hyperparameter for controlling smoothness of the distribution.
A lower temperature will assign most of the probability mass to a fewer documents while a higher temperature will make the distribution more uniform.
We employ cross entropy loss for $\mathrm{J}$, which is defined as
\begin{equation}
    \begin{aligned}
        \mathrm{J}(\mathbf{q}, \mathbf{d}, y; \boldsymbol{\theta}) &= -\log p(y|q, d; \boldsymbol{\theta}) \,\,\,\, \text{, where} \\
        p(y=1| q, d; \boldsymbol{\theta}) &= 
        \sigma (f_{\boldsymbol{\theta}} (\mathbf{q}, \mathbf{d}))\,\, , \,\,
        p(y=0|q, d; \boldsymbol{\theta}) = 
        1-\sigma (f_{\boldsymbol{\theta}} (\mathbf{q}, \mathbf{d}))\\
    \end{aligned}
\end{equation}
where $\sigma$ is a sigmoid function, and $f_{\boldsymbol{\theta}}$ is defined in Section \ref{sec:tasks} for each ad-hoc retrieval task.

The idea of adversarial sampling is not actually new.
Dynamic negative sampling ideas have been proposed by \cite{zhang2013optimizing,rendle2014improving}, which sample the most informational negative items at the moment.
A generator model of IRGAN \cite{wang2017irgan} also plays a similar role.
It builds the generator that can adversarially fool the discriminator, and it samples the negative examples according to the estimated generator.
Our framework is different from theirs in that we further generate more difficult adversarial negative examples from the adversarially sampled negative examples, and in that we generate adversarial examples even for positive class for better generalization.

\paragraph{Selective Virtual Adversarial Training}
Although the advantage of virtual adversarial training is that it can be used for all unlabeled documents, which is good for semi-supervised learning, it may not be effective and efficient to process many non-informational unlabeled documents.
Therefore, we propose selective virtual adversarial training based on adversarial sampling.
Instead of generating virtual adversarial perturbation for all unlabeled documents, we selectively generate them for difficult ones.
The objective is defined as
\begin{equation}
    \begin{aligned}
        \sum_q \Big( & \mathbb{E}_{d \sim p_{\text{data}}(d|q,y=1)} \big[ \mathrm{J}(\mathbf{q}, \mathbf{d}, y=1; \boldsymbol{\theta}) +
            \mathrm{J_{KL}}(\mathbf{q}, \mathbf{d}; \boldsymbol{\theta}) \big] + \\
        & \mathbb{E}_{d \sim p_{\boldsymbol{\theta}}(d|q,y=1) } \big[ \mathrm{J}(\mathbf{q}, \mathbf{d}, y=0; \boldsymbol{\theta}) +
            \mathrm{J_{KL}}(\mathbf{q}, \mathbf{d}; \boldsymbol{\theta}) \big]
        \Big)
    \end{aligned}
\end{equation}
where $\mathrm{J_{KL}}$ is defined in equation \eqref{eq:vat}.
Similar to equation \eqref{eq:adv_training}, we sample one negative document from unlabeled data by $p_{\boldsymbol{\theta}}$ for each positive document.
This is more efficient than the original virtual adversarial training \cite{miyato2015distributional} as it does not iterate over all unlabeled documents.
In addition, it can also be more effective because only highly informational documents, which may result in a better decision boundary, are selected to train the model.
Indeed, we empirically find that this approach is more effective than the original one in Section \ref{sec:result}.

\subsection{Pairwise Training}
Instead of collecting relevance level for each query-document pair, it is often easier for ad-hoc retrieval systems to infer which documents are more preferred than other documents for a query, leveraging implicit feedback of users.
In addition, users have biases, so one user's rating is often not compatible with another user, for example, in item recommendation.
Hence, it has been studied how to exploit such relative preference instead of absolute relevance with pairwise or listwise learning algorithms \cite{liu2009learning, rendle2009bpr}.
We thus extend our approaches to pairwise training.

There are several ways to form document pairs for pair-wise training.
For example, for a query $q$, we can build document pairs such that $\{ (d_i, d_j) | d_i \in D_{c} \wedge d_j \in D_{nc} \}$ where $D_c$ is a set of documents clicked by users and $D_{nc}$ is a set of documents that are not clicked.
There exist better ways to form such pairs, but they are not within the scope of this work.
The pairwise objective corresponding to equation \eqref{eq:adv_training} can be defined as
\begin{equation}\label{eq:pairwise_training}
    \begin{aligned}
        \sum_q & \Big( \mathbb{E}_{d^+ \sim p_{\text{data}}(d|q,y=1)} \Big[
        \mathbb{E}_{d^- \sim p_{\boldsymbol{\theta}}(d|q,y=1) } \big[ \\
        & \mathrm{J}(\mathbf{q}, \mathbf{d}^+, \mathbf{d}^-; \boldsymbol{\theta}) + 
        \mathrm{J}(\mathbf{q}+\boldsymbol{\eta}_q, \mathbf{d}^+ +\boldsymbol{\eta}_{d^+},\mathbf{d}^- +\boldsymbol{\eta}_{d^-}; \boldsymbol{\theta})
        \big] \Big] \Big)
    \end{aligned}
\end{equation}
where a preferred document $d^+$ is sampled from labeled data and the corresponding non-preferred document $d^-$ is sampled from unlabeled data according to $p_{\boldsymbol{\theta}}$.
The adversarial perturbations $\boldsymbol{\eta}_q$, $\boldsymbol{\eta}_{d^+}$, and  $\boldsymbol{\eta}_{d^-}$ can be computed in the same way as in equation \eqref{eq:general_taylor}.
The cost function $\mathrm{J}$ for pairwise training is defined as
\begin{equation}\label{eq:pairwise_loss}
    \begin{aligned}
        \mathrm{J}(\mathbf{q}, \mathbf{d}^+, \mathbf{d}^-; \boldsymbol{\theta}) & = 
        -\log p(y^+ \rhd y^- |q, d^+, d^-; \boldsymbol{\theta})
         \,\,\,\, \text{where} \\
         p(y^+ \rhd y^- |q, d^+, d^-; \boldsymbol{\theta}) & = 
        \sigma (f_{\boldsymbol{\theta}} (\mathbf{q}, \mathbf{d}^+) - f_{\boldsymbol{\theta}} (\mathbf{q}, \mathbf{d}^-) )
    \end{aligned}
\end{equation}
where $y^+ \rhd y^-$ means $d^+$ is preferred over $d^-$.

As in our solution, the pairwise perturbation is proposed for pairwise training as it assumes relative relevance between document pairs.
However, virtual adversarial perturbation assumes no labels, so pointwise learning makes more sense than pairwise learning.
Hence, we employ a hybrid learning whose objective is defined as
\begin{equation}\label{eq:pairwise_training_svat}
    \begin{aligned}
        \sum_q & \Big( \mathbb{E}_{d^+ \sim p_{\text{data}}(d|q,y=1)} \Big[
        \mathbb{E}_{d^- \sim p_{\boldsymbol{\theta}}(d|q,y=1) } \big[ \\
        & \mathrm{J}(\mathbf{q}, \mathbf{d}^+, \mathbf{d}^-; \boldsymbol{\theta}) + 
        \mathrm{J_{KL}}(\mathbf{q}, \mathbf{d}^+; \boldsymbol{\theta}) + 
        \mathrm{J_{KL}}(\mathbf{q}, \mathbf{d}^-; \boldsymbol{\theta})
        \big] \Big] \Big)
    \end{aligned}
\end{equation}
where we sample one negative example from unlabeled data by $p_{\boldsymbol{\theta}}$ for each positive example in the labeled data.
Here, we still learn from the pairwise loss with equation \eqref{eq:pairwise_loss} while we also learn from pointwise virtual adversarial loss (equation \eqref{eq:vat}) for each of positive and negative examples.

\section{Application to Ad-hoc Retrieval Tasks}\label{sec:tasks}
We apply the proposed approaches to ad-hoc retrieval tasks.
Among various ad-hoc retrieval tasks, we chose three tasks: Web search, item recommendation, and question answering.
These three tasks are specifically chosen because they are popular and actively studied, and they were also chosen by IRGAN \cite{wang2017irgan}, to which we mainly compare our approaches.

IRGAN is chosen as the main baseline because (\texttt{i}) it is based on a type of adversarial training too and it applies its unique negative sampling, (\texttt{ii}) it recently attracted a great deal of attention at ACM SIGIR 2017\footnote{\url{http://sigir.org/sigir2017/program/awards/}} with its novel approach and outstanding performance, and (\texttt{iii}) the source code and the pre-processed data sets are published\footnote{Available at \url{https://github.com/geek-ai/irgan} except Netflix data for item recommendation.} so that it provides an excellent benchmark environment.
For a fair comparison, we apply our approaches to the same models as IRGAN does, except for a small change in Web search model.
As the tasks and models are already described well in \cite{wang2017irgan}, we briefly describe them in this section. 

\subsection{Web Search}
For Web search task, input vectors for ($q$,$d$) can be formed as a length-$k$ vector $\mathbf{x}_{q,d} \in \mathbb{R}^k$, where each dimension represents a feature value of $q$ and/or $d$.
For example, TF-IDF scores of each document part and a query or PageRank scores of $d$ can be such features.
As in RankNet \cite{burges2005learning} and IRGAN\cite{wang2017irgan}, we adopt a two-layer neural network as the model of Web search:
\begin{equation}\label{eq:model_ir}
    \begin{aligned}
        f_{\boldsymbol{\theta}} (\mathbf{x}_{q,d}) = \mathbf{w}_2^\intercal f_1 (\mathbf{W}_1 \mathbf{x}_{q,d} + \mathbf{b}_1) + b_2
    \end{aligned}
\end{equation}
where $\mathbf{W}_1 \in \mathbb{R}^{l\times k}$ is a weight matrix and $l$ is the number of nodes in the hidden layer, $\mathbf{b}_1$ and $b_2$ are a bias vector and a constant, respectively, and $\mathbf{w}_2 \in \mathbb{R}^l$ is a weight vector for the output layer.
For the activation function in $f_1$, the hyperbolic tangent was employed in RankNet and IRGAN, but we replace it with ReLU since it can learn more effectively  \cite{glorot2011deep}.
Although ReLU is quite linear so that it is more vulnerable to adversarial examples, adversarial training makes the model robust to them, so it is not a concern.
We also experiment with ReLU version of IRGAN and compare the results.
Our approach is originally defined for two separate input vectors $\mathbf{q}, \mathbf{d}$ of $f_{\boldsymbol{\theta}}$, but accommodating a single input vector instead is straightforward. 

\subsection{Item Recommendation}
Item recommendation is a task where a ranked list of items are recommended for a user.
Thus, it can be regarded as an ad-hoc retrieval task, where a query is a user, and a document is an item.
One simple but popular approach is matrix factorization \cite{koren2009matrix} for collaborative filtering.
Given one-hot encoded vectors  $\mathbf{u}$ and $\mathbf{i}$ for a user $u$ and an item $i$, respectively, the model is defined as
\begin{equation}\label{eq:model_cf}
    \begin{aligned}
        f_{\boldsymbol{\theta}} (\mathbf{u}, \mathbf{i}) = \mathbf{v}_u^\intercal \mathbf{v}_i + b_i 
    \end{aligned}
\end{equation}
where $\mathbf{v}_u, \mathbf{v}_i \in \mathbb{R}^k$ are latent vectors for $u$ and $i$, respectively, and $b_i$ is a bias for $i$.
This model works in a linear way so that it is vulnerable to adversarial perturbation, so adversarial training is desired.

\subsection{Question Answering}
A retrieval-based question answering is a task where a ranked list of answers are retrieved for a question.
We employ a deep learning-based end-to-end approach to estimate latent vectors $\mathbf{v}_q, \mathbf{v}_d \in \mathbb{R}^k$.
Given one-hot encoded vectors $\mathbf{q}, \mathbf{d}$ for a question $q$ and an answer $d$, respectively, the relevance of $d$ to $q$ is modeled by a cosine similarity between them:
\begin{equation}\label{eq:model_qa}
    \begin{aligned}
        f_{\boldsymbol{\theta}} (\mathbf{q}, \mathbf{d}) = 
        \frac{ \mathbf{v}_q^\intercal \mathbf{v}_d }
        { ||\mathbf{v}_q|| \, ||\mathbf{v}_d|| }.
    \end{aligned}
\end{equation}
Recent successful approaches adopt convolutional neural network (CNN) \cite{severyn2016modeling,santos2016attentive} or long short-term memory neural network (LSTM) \cite{wang2015long} to obtain the latent vectors.
Although such models can achieve great performances, they are vulnerable to adversarial perturbation as they can behave in a linear way.
We thus apply our adversarial training to this model.

\section{Experiments on Three Ad-hoc Retrieval Tasks}\label{sec:result}

We perform experiments with our pairwise approaches as they often perform better than pointwise approaches.
The main approach uses the objective in equation \eqref{eq:pairwise_training}, which is based on pairwise adversarial training, and we call it \textbf{AdvIR} (Adversarial training for Information Retrieval).
We also experiment with selective virtual adversarial training (\textbf{AdvIR-SVAT}) in  equation \eqref{eq:pairwise_training_svat} and virtual adversarial training (\textbf{AdvIR-VAT}), which iterates over all unlabeled examples.
The code is available online.\footnote{\url{https://sites.google.com/site/daehpark/Resources}}

Due to the complexity of experiments, we borrow some baseline results from \cite{wang2017irgan}.
In fact, the exact pre-processed data set is published by \cite{wang2017irgan}, so we regard it as a concrete benchmark data set, and we do not repeat experiments for some baselines that are clearly inferior to IRGAN at least on those data sets.
Indeed, IRGAN is far superior to the baselines for Web search task.
The strongest competitors for other tasks, which are LambdaFM \cite{yuan2016lambdafm} for item recommendation and LambdaCNN \cite{zhang2013optimizing} for question answering, were actually proposed by some authors of IRGAN, so we can trust their results in \cite{wang2017irgan}.
For other settings, we follow the settings in \cite{wang2017irgan}.

\subsection{Web Search}

\subsubsection{Experimental Design}
The well-known benchmark data set LETOR 4.0 \cite{qin2013introducing} provides MQ2008-semi (Million Query track), which is designed for semi-supervised learning.
That is, it consists of a relatively small number of labeled data and a much larger number of unlabelled data.
Specifically, a relevance in 4 levels (-1, 0, 1, or 2) is given to each query-document pair, where -1 means unlabeled and a greater number means more relevance.
As we assume a single-class labeled data, we regard query-document pairs with relevance level 1 and 2 as `relevant' and compile labeled data, and compile unlabeled data with all other pairs.
As a result, there are 784 unique queries in total, and for each query, there are 5 labeled examples and 1,000 unlabeled examples on average.
For each query-document pair, there is a 46-dimension feature vector, which consists of continuous features such as TF-IDF and language model values.
The vector is given to the two-layer neural network in equation \eqref{eq:model_ir}.
The number of nodes in the hidden layer equals to the dimension size of the feature vector.
We set the perturbation size to 300 unless otherwise specified.

Popular learning to rank approaches including RankNet \cite{burges2005learning}, LambdaRank \cite{burges2007learning}, and LambdaMART \cite{burges2010ranknet} as well as IRGAN \cite{wang2017irgan} were employed as baselines.
Note that a generator of IRGAN plays a role of negative example sampler.
As our neural network employs ReLU \cite{glorot2011deep} for an activation function instead of hyperbolic tangent, we also train IRGAN with ReLU on both generator and discriminator, and we report their best measures.
We measure statistical significance of the improvement over this model with a paired t-test.
We adopt precision at $N$ and Normalized Discounted Cumulative Gain (NDCG) at $N$ as performance metrics since they are standard in ad-hoc retrieval.

\subsubsection{Result Analysis} 
    
\begin{table}[htb]
	\small
	\caption{ Overall results on Web search. The best value for each metric is bold-faced. $\ast$ indicates statistical significance over IRGAN\_ReLU. Relative improvement over IRGAN\_ReLU is shown in parentheses. } \label{tbl:ir_overall}
	\begin{tabular}{l l l l }
		\hline\hline
		 & \textbf{Prec@3} & \textbf{Prec@5} & \textbf{Prec@10} \\
		\hline
		RankNet \cite{burges2005learning} & 0.1619 & 0.1219 & 0.1010  \\
		LambdaRank \cite{burges2007learning} & 0.1651 & 0.1352 & 0.1076  \\
		LambdaMART \cite{burges2010ranknet} & 0.1368 & 0.1026 & 0.0846  \\
		IRGAN \cite{wang2017irgan} & 0.2000 & 0.1676 & 0.1248  \\
		IRGAN\_ReLU & 0.1937 & 0.1581 & 0.1286 \\
		\hline
		AdvIR & \textbf{0.2349}$^\ast$ (21.3\%) & 0.1829$^\ast$ (15.7\%) & 0.\textbf{1305} (1.5\%)  \\
		AdvIR\_VAT & 0.2000 (3.3\%) & 0.\textbf{1867}$^\ast$ (18.1\%)& 0.1248 (-3.0\%) \\
		AdvIR\_SVAT & 0.2222 (14.7\%) & 0.1810$^\ast$ (14.5\%) & 0.1238 (-3.7\%) \\
		\hline
		& \textbf{NDCG@3} & \textbf{NDCG@5} & \textbf{NDCG@10} \\
		\hline
		RankNet \cite{burges2005learning}  & 0.1801 & 0.1709 & 0.1943 \\
		LambdaRank \cite{burges2007learning}  & 0.1926 & 0.1920 & 0.2093 \\
		LambdaMART \cite{burges2010ranknet} & 0.1573 & 0.1456 & 0.1627 \\
		IRGAN \cite{wang2017irgan} &  0.2148 & 0.2154 & 0.2380 \\
		IRGAN\_ReLU &  0.2230 & 0.2185 & 0.2473 \\
		\hline
		AdvIR & 0.\textbf{2682}$^\ast$ (20.3\%) & \textbf{0.2568}$^\ast$ (17.5\%) & \textbf{0.2696}$^\ast$ (9.0\%)\\
		AdvIR\_VAT & 0.2390 (7.2\%)& 0.2512$^\ast$ (15.0\%)  & 0.2527 (2.2\%) \\
		AdvIR\_SVAT & 0.2598$^\ast$ (16.9\%) & 0.2538$^\ast$ (16.2\%) & 0.2593$^\ast$ (4.9\%)\\
		\hline\hline
	\end{tabular}
\end{table}

\begin{figure*}[bt] 
\centering
\begin{subfigure}{.25\textwidth}
    \centering
    \includegraphics[width=0.95\linewidth]{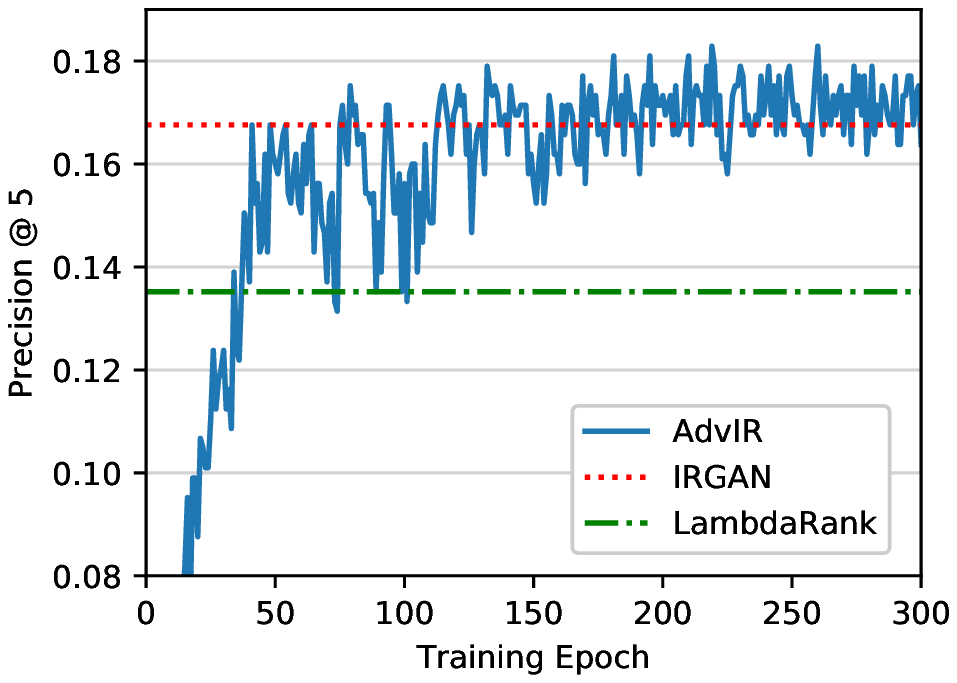}
    \caption{}
    \label{fig:ir_curve1}
\end{subfigure}%
\begin{subfigure}{.25\textwidth}
    \centering
    \includegraphics[width=0.95\linewidth]{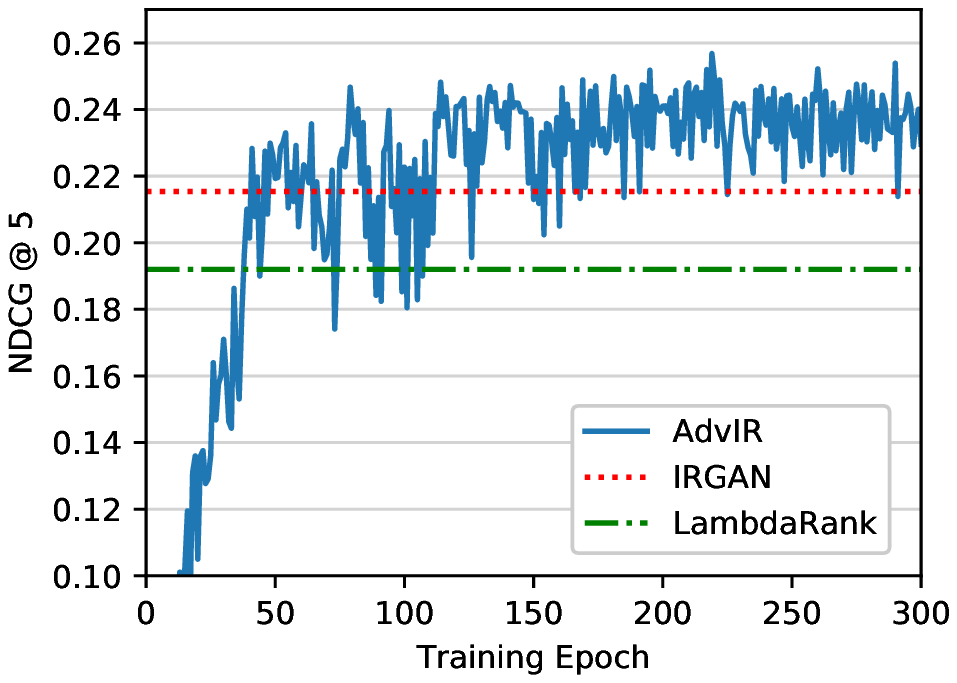}
    \caption{}
    \label{fig:ir_curve2}
\end{subfigure}%
\begin{subfigure}{.25\textwidth}
    \centering
    \includegraphics[width=0.95\linewidth]{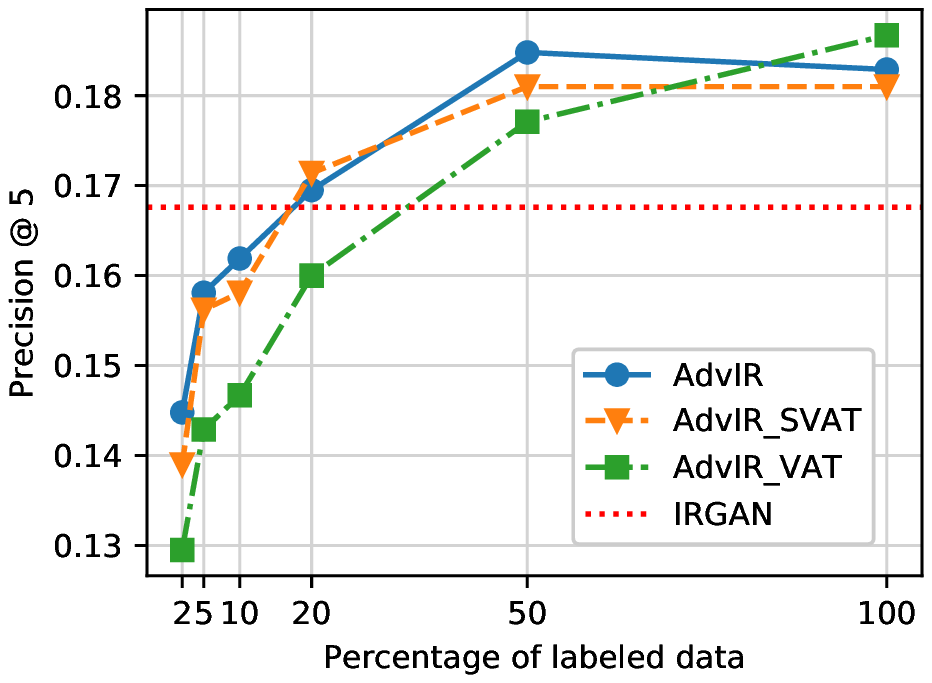}
    \caption{}
    \label{fig:ir_data1}
\end{subfigure}%
\begin{subfigure}{0.25\textwidth}
    \centering
    \includegraphics[width=0.95\linewidth]{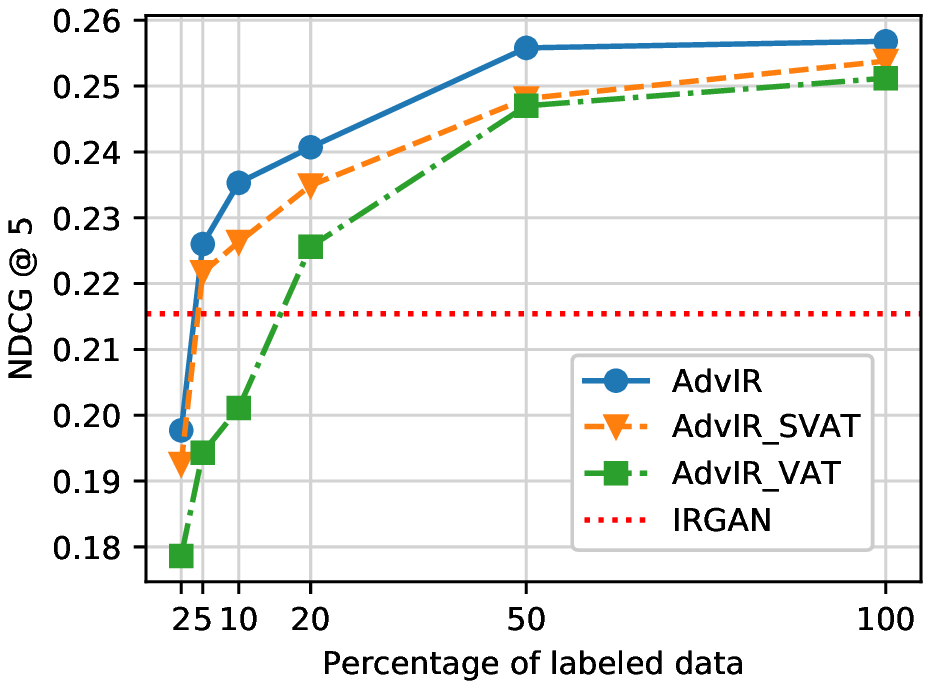}
    \caption{}
    \label{fig:ir_data2}
\end{subfigure}%
\caption{ (a,b) Learning curves and (c,d) data efficiency on Web search }
\label{fig:ir}
\end{figure*}

The overall results are shown in Table \ref{tbl:ir_overall}.
As described in \cite{wang2017irgan}, traditional learning to rank methods such as LambdaMART do not perform well since it is not specifically effective for semi-supervised learning.
IRGAN\_ReLU does not seem to differ much from IRGAN in terms of performance, but it performs better than IRGAN in NDCG measures. 
Our proposed approaches significantly outperform baselines in several measures.
The improvement from our approaches except AdvIR\_VAT is especially good for high-ranked documents.
This is expected because our approaches train models with the generated very difficult examples, so they are especially beneficial for distinguishing top documents, who are more difficult to rank in general.

Among the proposed methods, AdvIR outperforms the other proposed methods in general.
AdvIR\_SVAT outperforms AdvIR\_VAT especially for high-ranked documents, which can be identified by P@3 and N@3.
Similar to the previous analysis, this is reasonable because it focuses on difficult unlabeled data so that it can affect more in the high-ranked documents.
On the other hand, AdvIR\_VAT learns from all unlabeled data, and the too easy examples may affect the model in a negative way.

The learning curves of AdvIR are depicted in Figure \ref{fig:ir_curve1} and \ref{fig:ir_curve2}.
The dotted horizontal lines indicate the baseline methods' best measures among all training epochs.
We can see that from the early part of the training (epoch=50), it already outperforms the baselines on test data.
As more training is done, it easily outperforms the baselines.

In order to see how efficient the models are in terms of data, we randomly removed labeled data and recorded their performance in Figure \ref{fig:ir_data1} and \ref{fig:ir_data2}.
Surprisingly, two of our models, AdvIR and AdvIR\_SVAT can still outperform IRGAN in NDCG@5, using only 5\% of labeled data.
In Precision@5, they outperform IRGAN with only 20\% of labeled data.
Also, it is shown that AdvIR\_SVAT consistently outperforms AdvIR\_VAT especially when the number of labeled data is less.
Our conjecture is that as there are fewer labeled data, it is more important to focus on difficult unlabeled data to place the relevant documents in the top.

\begin{figure}[hbt] 
\centering
\includegraphics[width=0.6\linewidth]{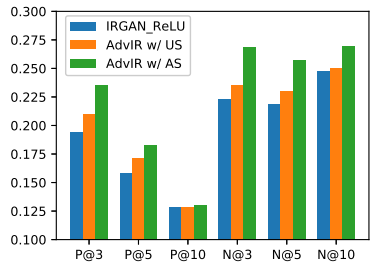}
\caption{ Effect of negative sampling methods for Web search. US stands for uniform sampling, and AS stands for adversarial sampling. }
\label{fig:ir_sampling}
\end{figure}

The effect of negative sampling methods on our proposed approach is depicted in Figure \ref{fig:ir_sampling}.
Even with uniform sampling for negative examples, AdvIR outperforms IRGAN, which has its own negative example sampling technique.
Our approach ``generates'' difficult negative examples instead of only ``sampling'' them, which is done by IRGAN.
Through the dynamically generated difficult examples, AdvIR can learn from diverse difficult examples.
In addition, IRGAN uses positive examples as they are whereas AdvIR generates difficult examples for even positive examples.
These differences can explain the superiority of AdvIR to IRGAN even when uniform sampling is used for AdvIR.
Switching from uniform sampling to adversarial sampling, AdvIR performs even better.
This is reasonable because the difficult negative examples sampled by adversarial sampling serve as great basis for generating even more difficult negative examples by adversarial training.

\subsection{Item Recommendation}
\subsubsection{Experimental Design}
To perform experiments for the item recommendation task, we apply our approaches on a popular benchmark data set, Movielens (100k) \cite{harper2016movielens}.
It consists of 943 users, 1,683 items, and 100,000 user-item ratings, where ratings are in 5 levels.
Following IRGAN \cite{wang2017irgan}, we regard the 4 and 5-star ratings as single-class labeled data\footnote{It is described in \cite{wang2017irgan} that only 5-star ratings are regarded as labeled data, but their published implementation regards both 4-star and 5-star ratings as labeled data.} and all other entries as unlabeled data.
We employ exactly the same data set as in \cite{wang2017irgan} where a 4:1 random splitting is done for training/test data.

The input vectors, which are one-hot encoded vectors, for users and items thus have length 943 and 1,683, respectively.
The input vectors are given to the matrix factorization model in equation \eqref{eq:model_cf}.
The size of latent vectors is 5.
We set the perturbation size to $0.01$ unless otherwise specified.

Popular and state-of-the-art baselines are employed, including Bayesian Personalised Ranking (BPR) \cite{rendle2009bpr}, LambdaRank-based collaborative filtering (LambdaFM) \cite{yuan2016lambdafm}, and IRGAN \cite{wang2017irgan}.
Note that BPR samples negative items from uniform distribution while LambdaFM and IRGAN dynamically sample difficult negative items.
To measure the statistical significance of our approach's improvement over IRGAN, we re-run IRGAN with their implementation and use the results to perform a paired t-test.
Standard performance metrics such as precision at $N$ and NDCG at $N$ are employed.

\subsubsection{Result Analysis}

\begin{table}[htb]
	\caption{ Overall results on item recommendation. $\ast$ indicates statistical significance over IRGAN\_rerun. } \label{tbl:cf_overall}
	\small
	\begin{tabular}{l l l l}
		\hline\hline
		 & \textbf{Prec@3} & \textbf{Prec@5} & \textbf{Prec@10} \\
		\hline
		BPR \cite{rendle2009bpr} & 0.3289 & 0.3044 & 0.2656  \\
		LambdaFM \cite{yuan2016lambdafm} & 0.3845 & 0.3474 & 0.2967  \\
		IRGAN \cite{wang2017irgan} & 0.4072 & 0.3750 & 0.3140  \\
		IRGAN\_rerun & 0.3999 & 0.3759 & 0.3217  \\
		\hline
		AdvIR & 0.4393$^\ast$ (9.9\%) & 0.4070$^\ast$ (8.3\%) & 0.3450$^\ast$ (7.2\%)  \\
		AdvIR\_VAT & 0.4313$^\ast$ (7.9\%) & \textbf{0.4083}$^\ast$ (8.6\%) & 0.3467$^\ast$ (7.8\%) \\
		AdvIR\_SVAT & \textbf{0.4466}$^\ast$ (11.7\%) & 0.4066$^\ast$ (8.2\%) & \textbf{0.3485}$^\ast$ (8.3\%) \\
		\hline
		& \textbf{NDCG@3} & \textbf{NDCG@5} & \textbf{NDCG@10} \\
		\hline
		BPR \cite{rendle2009bpr} &  0.3410 & 0.3245 & 0.3076 \\
		LambdaFM \cite{yuan2016lambdafm} &  0.3986 & 0.3749 & 0.3518 \\
		IRGAN \cite{wang2017irgan} &  0.4222 & 0.4009 & 0.3723 \\
		IRGAN\_rerun & 0.4166 & 0.4010 & 0.3779 \\
		\hline
		AdvIR &  0.4563$^\ast$ (9.5\%) & 0.4353$^\ast$ (8.6\%) & 0.4079$^\ast$ (7.9\%) \\
		AdvIR\_VAT & 0.4539$^\ast$ (8.9\%) & 0.4382$^\ast$ (9.3\%) & 0.4108$^\ast$ (8.7\%) \\
		AdvIR\_SVAT & \textbf{0.4641}$^\ast$ (11.4\%) & \textbf{0.4383}$^\ast$ (9.3\%) & \textbf{0.4183}$^\ast$ (10.7\%) \\
		\hline\hline
	\end{tabular}
	\normalsize
\end{table}

The overall results are shown in Table \ref{tbl:cf_overall}.
The results indicate the dynamic negative item sampling approaches (LambdaFM, IRGAN, and AdvIR) outperform BPR, showing that negative item sampling is indeed important.
Our proposed approaches significantly outperform baselines in all measures.
Similar to the Web search task, our proposed approaches perform especially well for high-ranked documents in general; the very difficult examples generated by them are supposed to be beneficial for ranking top documents.
Among our approaches, AdvIR\_SVAT performs the best for this task, but the difference is small.
Again, AdvIR and AdvIR\_SVAT performs especially better for high-ranked items (P@3 and N@3) than AdvIR\_VAT, with its focus on more difficult examples.

\begin{figure*}[hbt] 
\centering
\begin{subfigure}{.25\textwidth}
    \centering
    \includegraphics[width=0.95\linewidth]{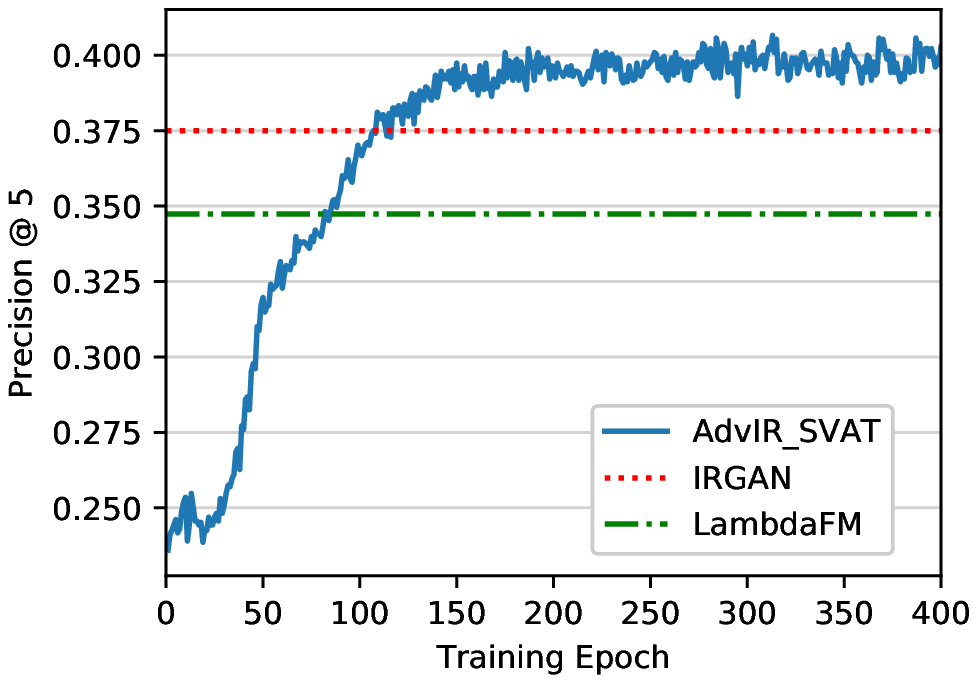}
    \caption{}
    \label{fig:cf_curve1}
\end{subfigure}%
\begin{subfigure}{.25\textwidth}
    \centering
    \includegraphics[width=0.95\linewidth]{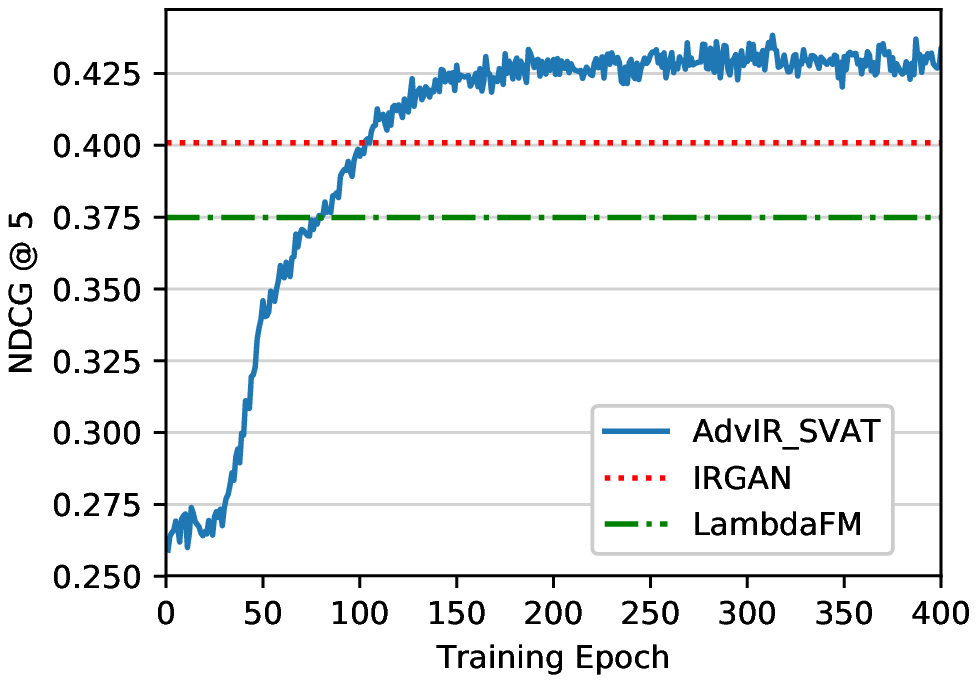}
    \caption{}
    \label{fig:cf_curve2}
\end{subfigure}%
\begin{subfigure}{.25\textwidth}
    \centering
    \includegraphics[width=0.95\linewidth]{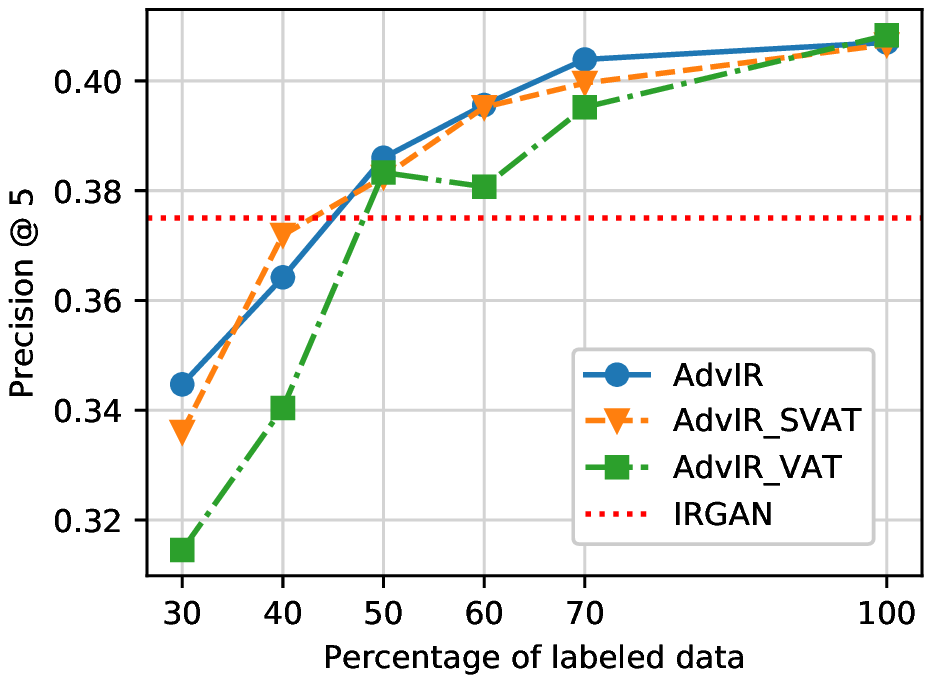}
    \caption{}
    \label{fig:cf_data1}
\end{subfigure}%
\begin{subfigure}{.25\textwidth}
    \centering
    \includegraphics[width=0.95\linewidth]{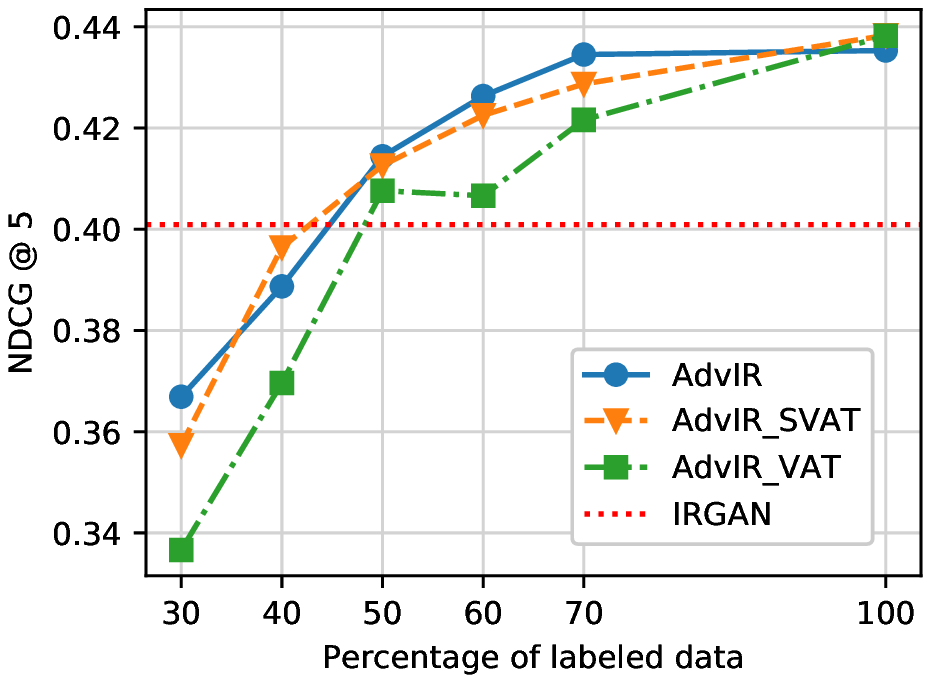}
    \caption{}
    \label{fig:cf_data2}
\end{subfigure}
\caption{ (a,b) Learning curves and (c,d) data efficiency on item recommendation   }
\label{fig:cf}
\end{figure*}

The learning curves of AdvIR\_SVAT are depicted in Figure \ref{fig:cf_curve1} and \ref{fig:cf_curve2}.
At around only 100 epochs, it starts to outperform the best measures of baselines.
As more training is done, it shows better performance on the test data, and it gains little after ~300 epochs.
Different from Web search, the curves do not oscillate much for item recommendation.

The data efficiency is depicted in Figure \ref{fig:cf_data1} and \ref{fig:cf_data2}.
Using only 50\% of the labeled data, our proposed models outperform IRGAN in both Precision@5 and NDCG@5.
Although our proposed approaches show significant improvements, its data efficiency for the item recommendation task is not as good as that for the Web search task.
It is reasonable because it is more difficult to perform well for collaborative filtering when the number of labeled data is less; the information in transactions (user-item ratings) directly shrinks so that the models suffer from the cold-start problem.
In other words, Web search can take advantage of text in the document when matching a document with a query, but a user does not have such information for collaborative filtering but has only user-item ratings; hence, it is much more difficult to perform well for item recommendation if there are fewer labeled data (user-item ratings).
Similar to Web search results, it is shown that AdvIR\_SVAT consistently outperforms AdvIR\_VAT especially for fewer labeled data, with its focus on difficult unlabeled examples.

\begin{figure}[hbt] 
\centering
\includegraphics[width=0.6\linewidth]{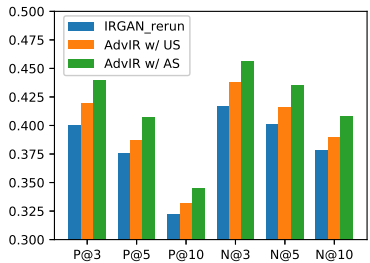}
\caption{ Effect of negative sampling methods for item recommendation. US stands for uniform sampling, and AS stands for adversarial sampling. }
\label{fig:cf_sampling}
\end{figure}

The effect of negative sampling methods is depicted in Figure \ref{fig:cf_sampling}.
Similar to Web search results, AdvIR with uniform negative sampling still outperforms IRGAN, which has its own dynamic negative sampling mechanism.
IRGAN only samples negative examples while AdvIR generates them on top of sampled negative examples.
AdvIR also generates difficult examples for positive examples while IRGAN does not.
These may have caused the superiority of AdvIR to IRGAN.
Adversarial sampling again seems to play an important role in AdvIR for item recommendation.
Adversarial sampling helps AdvIR to generate more difficult examples compared to uniform sampling, so AdvIR with adversarial sampling consistently outperforms that with uniform sampling.

\subsection{Question Answering} 
\subsubsection{Experimental Design}
For question answering, we apply our approaches to one of the popular benchmark data sets, Insurance QA \cite{feng2015applying}.
It consists of questions, which are submitted by users, and high-quality answers, which are written by domain experts.
Training data consist of 12,887 (question, answer) pairs, and development data consist of 1,000 (question, answer) pairs.
That is, a single-class labeled data (answer) is given for the task.
We can thus regard the whole answer sets as unlabeled data to apply semi-supervised learning.
There are two test sets (\texttt{test-1} and \texttt{test-2}), each of which consists of 1,800 (question, answer) pairs.
The task is to retrieve the one and only correct answer from 500 candidate answers.
Hence, we report precision @ 1 as the performance metric.

Both questions and answers are in raw text, so we apply the end-to-end model based on convolutional neural networks (CNN).
The input vectors, which are one-hot encoded vectors, for questions and answers thus have length that is the vocabulary size, and the embedding vectors have 100 dimensions.
We use the same CNN architecture as in IRGAN.
The convolutional layer consists of 4 different kernels with sizes 1, 2, 3, and 5, and the feature maps after applying those kernels to the embedding are summarized by the max-pooling-over-time.
Then, the resulting vectors for a question and an answer are given to equation \eqref{eq:model_qa} as $\mathbf{v}_q$ and $\mathbf{v}_d$, respectively.
We set the perturbation size to 0.5 unless otherwise specified.

Strong baselines such as QA-CNN \cite{santos2016attentive}, LambdaCNN \cite{santos2016attentive, zhang2013optimizing} that enhances QA-CNN with dynamic negative sampling, and IRGAN \cite{wang2017irgan} are employed.
Note that uniform negative sampling is done by QA-CNN, and dynamic negative sampling is done in LambdaCNN and IRGAN. 
To measure the statistical significance of our approaches' improvement over IRGAN, we re-run IRGAN with their implementation and use the results to perform a paired t-test.

\subsubsection{Result Analysis}

\begin{table}[htb]
	\caption{ Overall results on question answering. Precision@1 is reported. } \label{tbl:qa_overall}
	\begin{tabular}{l l l }
		\hline\hline
		 & \textbf{test-1} & \textbf{test-2} \\
		\hline
		QA-CNN \cite{santos2016attentive} & 0.6133 & 0.5689 \\
		LambdaCNN \cite{santos2016attentive,zhang2013optimizing} & 0.6294 & 0.6006 \\
		IRGAN \cite{wang2017irgan} & 0.6444 & 0.6111 \\
		IRGAN\_rerun & 0.6478 & 0.6028 \\
		\hline
		AdvIR & 0.6489 (0.2\%) & \textbf{0.6150} (2.0\%)\\
		AdvIR\_VAT & 0.6450 (-0.4\%) & 0.6083 (0.9\%) \\
		AdvIR\_SVAT & \textbf{0.6517} (0.6\%) & 0.6133 (1.7\%) \\
		\hline\hline
	\end{tabular}
	\normalsize
\end{table}

The overall results on question answering are shown in Table \ref{tbl:qa_overall}.
Approaches based on dynamic negative sampling techniques, including our approaches, IRGAN, and Lambda CNN, outperform QA-CNN that is based on uniform negative sampling, but the difference in performance is not as big as that of Web search or item recommendation tasks.
Although our proposed approaches outperform IRGAN in most measures, the improvements are not statistically significant.
The performance difference among our proposed approaches is also small.
There can be a couple of reasons for these.
As the negative answers in the test data come from answers of other questions, the task is easier than other tasks.
Thus, the advantage comes from learning with difficult examples may not be clearly shown  in this data set.
Indeed, the already high measures in Precision@1 tells the task is relatively easy.
Also, there may not be negative examples that are difficult enough due to the data setting.
A small improvement by dynamic negative sampling approaches compared to other tasks supports this.

\begin{figure}[hbt] 
\centering
\begin{subfigure}{.5\columnwidth}
    \centering
    \includegraphics[width=0.95\linewidth]{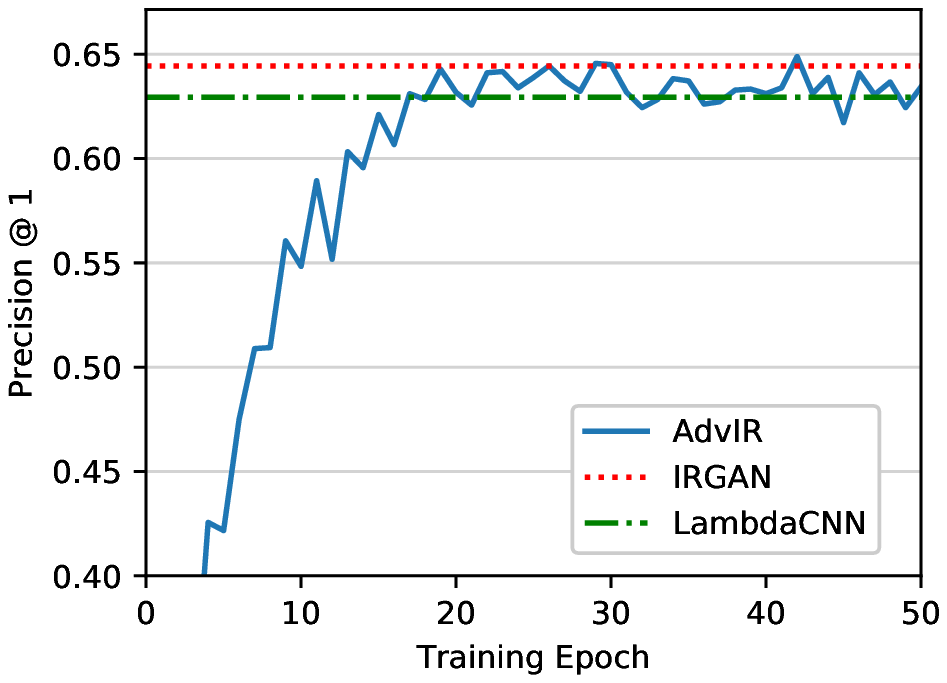}
    \label{fig:accuracy_drop_1}
\end{subfigure}%
\begin{subfigure}{.5\columnwidth}
    \centering
    \includegraphics[width=0.95\linewidth]{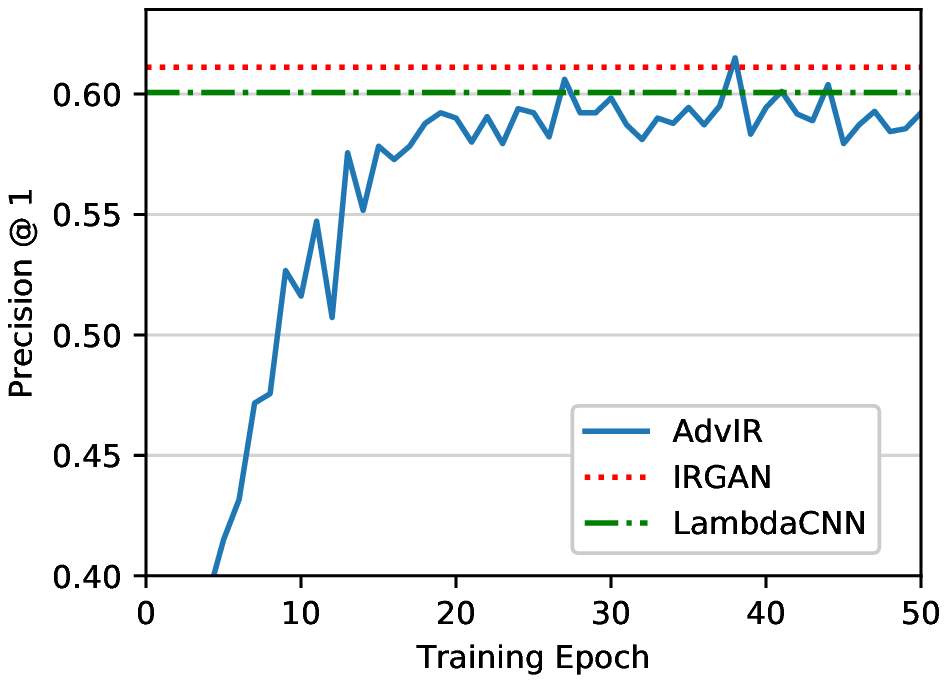}
    \label{fig:accuracy_drop_2}
\end{subfigure}
\caption{ Learning curves on question answering for \texttt{test-1} (left) and \texttt{test-2} (right).  }
\label{fig:qa_curve}
\end{figure}

The learning curves of AdvIR are depicted in Figure \ref{fig:qa_curve}.
Although it does not take advantage of pre-trained model as in IRGAN, AdvIR reaches high precision quickly (in 20 epochs).
On the other hand, IRGAN, which starts with Precision@1 of 0.6 by pre-trained model, takes >30 epochs to reach Precision@1 >= 0.64 \cite{wang2017irgan}.

\section{Conclusions}

In this paper, we studied semi-supervised ad-hoc retrieval models with implicit feedback, where there are relatively fewer single-class labeled data and much more unlabeled data.
We proposed an adversarial sampling and training framework that handles labeled data and unlabeled data differently.
It adversarially generates informational examples for the positive class; on the other hand, it first adversarially samples informational examples for the negative class and further adversarially generates even more informational negative examples.
We also proposed virtual adversarial training and selective virtual adversarial training that is more effective and efficient than the former, and they do not require labels to generate adversarial examples.

Experiments were performed on public benchmark data sets for three popular ad-hoc retrieval tasks such as Web search, item recommendation, and question answering.
Experimental results show that (\texttt{i}) our proposed approaches are effective on ad-hoc retrieval tasks: they significantly outperform baselines on Web search and item recommendation and are on par with IRGAN on question answering, (\texttt{ii}) our proposed approaches perform especially well for high-ranked documents, (\texttt{iii}) our proposed approaches are data-efficient; (\texttt{iv}) adversarial sampling amplifies the effectiveness of adversarial training, and (\texttt{v}) our proposed selective virtual adversarial training is more effective and efficient than virtual adversarial training.

\bibliographystyle{ACM-Reference-Format}
\balance
\bibliography{bib}

\end{document}